\documentclass[aps,prd,12pt,amssymb]{revtex4}

\begin{document}

\baselineskip=0.60cm

\newcommand{\ini}{\begin{equation}}
\newcommand{\fin}{\end{equation}}
\newcommand{\inir}{\begin{eqnarray}}
\newcommand{\finr}{\end{eqnarray}}
\newcommand{\inif}{\begin{figure}}
\newcommand{\finf}{\end{figure}}
\newcommand{\bc}{\begin{center}}
\newcommand{\ec}{\end{center}}

\def\ol{\overline}
\def\pa{\partial}
\def\ra{\rightarrow}
\def\ts{\times}
\def\df{\dotfill}
\def\bs{\backslash}
\def\dg{\dagger}

$~$

\hfill DSF-17/2003

\vspace{1 cm}

\title{From neutrino oscillations to baryogenesis
\footnote{Based on the talk presented at the Fourth International
School of Physics ``Bruno Pontecorvo", Capri, May 26-29, 2003}}

\author{D. Falcone}

\affiliation{Dipartimento di Scienze Fisiche,
Universit\`a di Napoli, Via Cintia, Napoli, Italy}

\begin{abstract}
\vspace{1cm}
\noindent
The evidence for neutrino oscillations leads to small neutrino masses,
which can be realized by means of the seesaw mechanism. In this framework,
baryogenesis may be achieved from leptogenesis.
\end{abstract}

\maketitle

\newpage

{\bf I. Brief history of neutrino oscillations}

$~$

The concept of neutrino oscillations was introduced by B. Pontecorvo in 1957
\cite{p1}. He considered $\nu-\ol{\nu}$ oscillations in vacuum, in analogy
to $K-\ol{K}$ oscillations \cite{gp}.

$~$

Flavor mixing was proposed by Maki, Nakagawa and Sakata in 1962 \cite{mns}.
According to this idea, two weak (flavor) eigenstates $\nu_{\alpha}$
are related to two mass eigenstates $\nu_i$ by a rotation $U$, that is
$\nu_{\alpha}=U_{\alpha i} \nu_i$. More generally, $U$ is a unitary matrix.

$~$

Flavor oscillations of Majorana neutrino were introduced by B. Pontecorvo
in 1967 \cite{p2}. In this paper he also anticipated the solar neutrino
problem, since he pointed out that, due to neutrino oscillations, the
observed flux of solar neutrinos should be half of the expected flux.
In fact, in 1968, a deficit of solar neutrinos ($\nu_e$) was found
\cite{dhh}, with respect to the theoretical calculation on the basis of the
solar model \cite{bbs}.
Then, in 1969, Gribov and Pontecorvo proposed the solution
of the solar neutrino problem by means of neutrino oscillations in vacuum
\cite{gb}. The pattern of oscillations is modified in matter \cite{msw}.
Several years later, the atmospheric neutrino anomaly was discovered
\cite{atm}, that is a deficit in atmospheric neutrinos ($\nu_{\mu}$).

$~$

Evidence for neutrino oscillations in atmospheric neutrinos was indeed
found in 1998 by the SuperKamiokande experiment \cite{sk}.
Then, in 2002, evidence for neutrino oscillations in solar neutrinos
has been also found \cite{sno}. Finally, in 2003, terrestrial evidence for
$\ol{\nu_e}$ oscillations from reactor neutrinos \cite{kamland}
and terrestrial evidence for $\nu_{\mu}$ oscillations from accelerator neutrinos
\cite{k2k} are achieved.

$~$

Hence, the solar neutrino problem is solved by $\nu_e-\nu_{\mu,\tau}$
matter oscillations in the sun, while the atmospheric neutrino anomaly
is explained by $\nu_{\mu}-\nu_{\tau}$ vacuum oscillations.

$~$

{\bf II. Neutrino masses and mixings}

$~$

The oscillation formula is given by the expression
\ini
P \simeq \sin^2 2 \theta \sin^2 \frac{\Delta m^2 L}{4 E},
\fin
where P is the probability of transition at distance $L$ from the source,
$E$ is the energy of neutrinos, $\theta$ is the mixing angle, and 
$\Delta m^2$ is the square mess difference between the two mass eigenstates
involved in the process. Therefore, neutrino oscillations imply neutrino
masses and mixings. From oscillation data, the following values are inferred:
\ini
\Delta m^2_{32} = m_3^2-m_2^2 \simeq 2.7 \cdot 10^{-3} \text{eV}^2,
\fin
\ini
\Delta m^2_{21} = m_2^2-m_1^2 \simeq 7.1 \cdot 10^{-5} \text{eV}^2,
\fin
where the three masses $m_1$, $m_2$, $m_3$ are the effective neutrino masses.
The lepton mixing matrix, is given by the approximate form
\ini
U \simeq \left( \begin{array}{ccc}
\frac{\sqrt2}{\sqrt3} & \frac{1}{\sqrt3} & 0 \\
-\frac{1}{\sqrt6} & \frac{1}{\sqrt3} & \frac{1}{\sqrt2} \\
\frac{1}{\sqrt6} & -\frac{1}{\sqrt3} & \frac{1}{\sqrt2}
\end{array} \right).
\fin
This expression yields large $U_{e2}$, near maximal $U_{\mu 3}$, and
small $U_{e3}$. Hence, lepton mixings can be large and even maximal,
while quark mixings are all small.

$~$

Neutrino masses are very small with respect to charged fermion masses.
In fact, from beta decay experiments we get
\ini
m_{\nu_e}=(U_{ei}^2 m_i^2)^{1/2} < 2.2 \text{eV},
\fin
which gives $m_i \lesssim 1$ eV.
Also from cosmology we get $\sum m_i \lesssim 1$ eV.
Moreover, for Majorana neutrinos, a upper limit comes from neutrinoless
double beta decay,
\ini
M_{ee}=(U_{ei}^2 m_i) < 0.86 \text{eV}.
\fin
In contrast, charged fermion masses
span the range going from $m_e \sim 1$ MeV to $m_t \sim 100$ GeV.
We should find a mechanism for generating very small neutrino masses.

$~$

{\bf III. Dirac and Majorana masses}

$~$

A Dirac mass term can be written in the form
\ini
m_D \ol{\psi_R} \psi_L +h.c..
\fin
It conserves electric charge and lepton number.

$~$

Majorana mass terms for left-handed and right-handed particles can be written as
\ini
m_L \ol{(\psi_L)^c} \psi_L +h.c.,
\fin
\ini
m_R \ol{(\psi_R)^c} \psi_R +h.c.,
\fin
respectively.
They violate electric charge and lepton number, thus are allowed only for
neutral particles, in particular neutrinos.

$~$

A Dirac neutrino is expressed by the field $(\nu_L,\nu_R)$, which under
charge conjugation goes into $((\nu_R)^c,(\nu_L)^c)$. Instead, Majorana
neutrino fields are given by $(\nu_L,(\nu_L)^c)$ and $((\nu_R)^c,\nu_R)$,
which are both self-conjugate.
Majorana fields contain only one type of Weyl spinor, left or right, while
Dirac fields contain both types of spinors.

$~$

{\bf IV. Seesaw mechanism}

$~$

If both Dirac and Majorana masses are present, the full mass matrix of
neutrinos is
\ini
M=\left( \begin{array}{cc}
m_L & m_D \\ m_D & m_R
\end{array} \right).
\fin
Let us assume that $m_L=0$, and $m_R \gg m_D$. Then $M$ has mass eigenvalues
nearly equal to $m_R$ and $m_D^2/m_R$. The latter can be written as
$(m_D/m_R) m_D$, so that it is much smaller than $m_D$. This is the
seesaw mechanism \cite{ss}. The corresponding two eigenstates are of the
Majorana kind, see for example \cite{kay}.
For three generations of fermions, we have the seesaw formula for mass matrices,
\ini
M_L \simeq M_{\nu} M_R^{-1} M_{\nu}^T,
\fin
where $M_L$ is the effective mass matrix of (light) left-handed neutrinos,
$M_{\nu}$ is the Dirac neutrino mass matrix, and $M_R$ is the mass matrix of
(heavy) right-handed neutrinos.

$~$

In the minimal standard model, the neutrino is massless because $\nu_R$
does not exist. The minimal extension is then adding $\nu_R$. Gauge extension,
such as the left-right model, the Pati-Salam model, and the $SO(10)$ unified
model, do include the right-handed neutrino. It is natural for the Dirac mass
to be similar to charged fermion masses, since all are generated from couplings
to the same Higgs fields. In particular, we expect $M_e \sim M_d$ and
$M_{\nu} \sim M_u$. Instead, the Majorana mass matrix of right-handed neutrinos
is generated as bare mass term or by coupling to another Higgs field, so that
its value can be very large. These conditions lead to the seesaw mechanism.

$~$

Moreover, a seesaw enhancement of lepton mixing can appear \cite{smirnov}.
For instance, let us take
\ini
M_{\nu} \simeq \left( \begin{array}{ccc}
0 & a & 0 \\ a & b & c \\
0 & c & 1
\end{array} \right) m_t,
\fin
with $a \ll b \sim c \ll 1$, and \cite{falc}
\ini
M_{R} \simeq \left( \begin{array}{ccc}
c^2 & 0 & 0 \\ 0 & c^2 & 0 \\ 0 & 0 & 1
\end{array} \right) m_R.
\fin
Then we obtain
\ini
M_L \simeq \left( \begin{array}{ccc}
k^2 & k & k \\ k & 1 & 1 \\ k & 1 & 1
\end{array} \right) \frac{m_t^2}{m_R},
\fin
with $k=a/c$. Other minimal forms which give large lepton mixing are
\cite{falc}
\ini
M_{R} \simeq \left( \begin{array}{ccc}
0 & ac & 0 \\ ac & 0 & 0 \\ 0 & 0 & 1
\end{array} \right) m_R,
\fin
\ini
M_{R} \simeq \left( \begin{array}{ccc}
\frac{a^2}{c} & 0 & 0 \\ 0 & 0 & 1 \\ 0 & 1 & 0
\end{array} \right) m_R.
\fin

$~$

In another approach, we may invert the seesaw formula,
\ini
M_R \simeq M_{\nu}^T M_L^{-1} M_{\nu},
\fin
in order to determine the heavy neutrino mass matrix.
Assuming $M_{\nu} \sim M_u$, that is $a \sim \lambda^6$, $c \sim \lambda^4$,
we get two possible forms
\cite{fal}
\ini
M_R \sim \left( \begin{array}{ccc}
\lambda^{12} & \lambda^{10} & \lambda^6 \\
\lambda^{10} & \lambda^8 & \lambda^4 \\
\lambda^6 & \lambda^4 & 1
\end{array} \right) \frac{m_t^2}{m_k},
\fin
with eigenvalues $M_1/M_2 \sim \lambda^4$, $M_1/M_3 \sim \lambda^{12}$, and
\ini
M_R \sim \left( \begin{array}{ccc}
0 & \lambda^6 & 0 \\
\lambda^6 & \lambda^4 & 1 \\
0 & 1 & 0
\end{array} \right) \lambda^6 \frac{m_t^2}{m_1},
\fin
with eigenvalues $M_1/M_2 \sim \lambda^6$, $M_1/M_3 \sim \lambda^6$.
The first form has small mixings, while the second one has large mixing
in the 2-3 sector.

$~$

{\bf V. Baryogenesis from leptogenesis}

$~$

It is interesting to see the implications of the previous matrix models
for the baryogenesis via leptogenesis mechanism \cite{fy}.
This is based on the out-of-equilibrium decays of
the right-handed neutrinos, which produce a lepton asymmetry, partially
converted to a baryon asymmetry by electroweak sphalerons. The baryon asymmetry
is given by
\ini
Y_B \simeq \frac{1}{2} \frac{1}{g^*} ~d ~\epsilon_1,
\fin
where $\epsilon_1$ can be written as
\ini
\epsilon_1 \simeq \frac{3}{16 \pi v^2}
\left[ \frac{(M_D^{\dg} M_D)_{12}^2}{(M_D^{\dg} M_D)_{11}} \frac{M_1}{M_2}+
\frac{(M_D^{\dg} M_D)_{13}^2}{(M_D^{\dg} M_D)_{11}} \frac{M_1}{M_3} \right].
\fin
Here, $M_D$ is the Dirac neutrino mass matrix in the basis where $M_R$ is
diagonal, and $M_i$ are the heavy neutrino masses. The parameter $d<1$ is a
dilution factor, and $g^* \simeq 100$. The allowed value for the baryon
asymmetry is
$Y_B \simeq 9 \cdot 10 ^{-11}$.

$~$

For matrix models (13) and (15), sufficient baryon asymmetry can be 
obtained, while matrix model (16) give too small asymmetry. 

$~$

For the two matrix models (18) and (19),
we get a too low baryon asymmetry \cite{fal}. Then, we proposed
another mass matrix model,
where the overall mass scale of $M_{\nu}$ is again $m_t$, but the internal
hierarchy is that of $M_d$ and $M_e$, namely $a \sim \lambda^3$,
$c \sim \lambda^2$. In this case a sufficient amount of
baryon asymmetry is achieved \cite{fal}.

$~$

For other recent studies of the
relation between seesaw mechanism and leptogenesis, see
Refs.\cite{ft,afs}. In particular, baryon asymmetry is enhanced for
$M_1 \sim M_2$, as happens for models (13) and (15).

$~$

{\bf VI. Conclusion}

$~$

It is quite impressive the chain which takes us from neutrino oscillations
to small neutrino masses, to the seesaw mechanism, and to baryogenesis
through leptogenesis. If the seesaw mechanism is indeed correct, we should
find confirmed evidence for the neutrinoless double beta decay.

$~$

The author thanks prof. F. Buccella for invitation to the School.


\begin{thebibliography}{100}

\newpage

\bibitem{p1} B. Pontecorvo, Zh. Eksp. Teor. Fiz. 33 (1957) 549 
[Sov. Phys. JETP 6 (1957) 429]

B. Pontecorvo, Zh. Eksp. Teor. Fiz. 34 (1957) 247
[Sov. Phys. JETP 7 (1958) 172]

\bibitem{gp} M. Gell-Mann and A. Pais, Phys. Rev. 97 (1955) 1387 

\bibitem{mns}  Z. Maki, M. Nakagawa and  S. Sakata,
Prog. Theor. Phys. 28 (1962) 870

\bibitem{p2} B. Pontecorvo, Zh. Eksp. Teor. Fiz. 53 (1967) 1717
[Sov. Phys. JETP 26 (1968) 984]

\bibitem{dhh} R. Davis, Jr., D.S. Harmer and K.C. Hoffman,
Phys. Rev. Lett. 20 (1968) 1205

Neutrinos are detected according to the reaction proposed in:

B. Pontecorvo, Chalk River Report PD-205 (1946)

\bibitem{bbs} J.N. Bahcall, N.A. Bahcall and G. Shaviv,
Phys. Rev. Lett. 20 (1968) 1209  

\bibitem{gb} V.N. Gribov and B. Pontecorvo, Phys. Lett. B 28 (1969) 493 

\bibitem{msw} L. Wolfenstein, Phys. Rev. D 17 (1978) 2369;
S.P. Mikheev and A.Yu. Smirnov, Yad. Fiz. 42 (1985) 1441 
[Sov. J. Nucl. Phys. 42 (1985) 913]

\bibitem{atm} T.J. Haines et al., Phys. Rev. Lett. 57 (1986) 1986

K.S. Hirata et al. (Kamiokande Collaboration), Phys. Lett. B 205 (1988)
416 

D. Casper et al.,  Phys. Rev. Lett. 66 (1991) 2561  

\bibitem{sk} Y. Fukuda et al. (Super-Kamiokande Collaboration),
Phys. Rev. Lett. 81 (1998) 1562

\bibitem{sno} Q.R. Ahmad et al. (SNO Collaboration),
Phys. Rev. Lett. 89 (2002) 011301 

\bibitem{kamland} K. Eguchi et al. (KamLAND Collaboration),
Phys. Rev. Lett. 90 (2003) 021802

\bibitem{k2k} M.H. Ahn et al. (K2K Collaboration),
Phys. Rev. Lett. 90 (2003) 041801

\bibitem{ss} M. Gell-Mann, P. Ramond and R. Slansky, in
$Supergravity$, eds. P. van Nieuwenhuizen and D. Freedman
(North Holland, Amsterdam, 1979)

T.Yanagida, in $Proceedings$ $of~the$ $Workshop$ $on$ $Unified$ $Theories$
$and$ $Baryon$ $Number$ $in~the$ $Universe$, eds. O. Sawada and A. Sugamoto
(KEK, Tsukuba, 1979)

S.L. Glashow, in $Quarks~and~Leptons$, eds. M. Levy et al. (Plenum, New York,
1980)

R.N. Mohapatra and G. Senjanovic, Phys. Rev. Lett. 44 (1980) 912

\bibitem{kay} B. Kayser, hep-ph/0211134

\bibitem{smirnov} A.Yu. Smirnov, Phys. Rev. D 48 (1993) 3264

\bibitem{falc} D. Falcone, hep-ph/0303074

\bibitem{fal} D. Falcone, hep-ph/0305229

\bibitem{fy} M. Fukugita and T. Yanagida, Phys. Lett. B 174 (1986) 45 

\bibitem{ft} D. Falcone and F. Tramontano, Phys. Rev. D 63 (2001) 073007

E. Nezri and J. Orloff, J. High Energy Phys. 04 (2003) 020 

\bibitem{afs} E. Kh. Akhmedov, M. Frigerio and A.Yu. Smirnov, hep-ph/0305322

\end{thebibliography}
\end{document}